\documentclass[prl,aps,twocolumn,superscriptaddress]{revtex4}
\usepackage[T1]{fontenc}

\usepackage[utf8]{inputenc}
\usepackage{amsmath,amssymb}
\usepackage{amsfonts}
\usepackage{epsfig,pstricks,graphicx}
\usepackage{bm}
\usepackage{color}
\def\id{{\rm 1\kern-.22em l}}
\usepackage{enumerate}

\begin{document}

\title{Demonstration of quantum controlled teleportation for discrete variables on linear optical devices} 
\author{Artur Barasi\'nski}
\email{artur.barasinski@upol.cz}
\affiliation{RCPTM, Joint Laboratory of Optics of Palack\'{y} University and Institute of Physics of CAS, Faculty of Science, Palack\'{y} University, 17. Listopadu 12, 771 46 Olomouc, Czech Republic}
\affiliation{Institute of Physics, University of Zielona G\'ora, Z. Szafrana 4a, 65-516 Zielona G\' ora, Poland}

\author{Anton\'{i}n \v{C}ernoch}
\email{antonin.cernoch@upol.cz}
\affiliation{RCPTM, Joint Laboratory of Optics of Palack\'{y} University and Institute of Physics of CAS, Faculty of Science, Palack\'{y} University, 17. Listopadu 12, 771 46 Olomouc, Czech Republic}

\author{Karel Lemr}
\email{k.lemr@upol.cz}
\affiliation{RCPTM, Joint Laboratory of Optics of Palack\'{y} University and Institute of Physics of CAS, Faculty of Science, Palack\'{y} University, 17. Listopadu 12, 771 46 Olomouc, Czech Republic}

\begin{abstract}
We report an experimental implementation of tripartite controlled quantum teleportation on the quantum optical devices. 
The protocol is performed through bi- and tripartite entangled channels of discrete variables and qubits encoded in polarization of individual photons. 
The experimental results demonstrate successful controlled quantum teleportation with a fidelity around $83\%$, well above the classical limit.
By realizing the controlled quantum teleportation through biseparable state, we show that tripartite entangled is not a necessary resource for controlled quantum teleportation and the controller's capability to allow or prohibit the teleportation cannot be considered to be a manifestation of tripartite entanglement. 
These results open new possibilities for further application of controlled quantum teleportation by lowering teleportation channel's requirements.
\end{abstract}

\maketitle

%%%%%%%%%%%%%%%%%%%%%%%%%%%%%%%%%%%%%%%%%%%%%%%%%%%%%%%%%%%%%%%%%%%
%\section{Introduction}
%%%%%%%%%%%%%%%%%%%%%%%%%%%%%%%%%%%%%%%%%%%%%%%%%%%%%%%%%%%%%%%%%%%

{\em Introduction.---}Quantum teleportation is considered as one of the major protocols in quantum information science. By exploiting the physical resource of entanglement, quantum teleportation has played a prominent role in the development of quantum information theory \cite{Bennettprl70_1993,Nielsencambridge_2000,Kimprl86_2001,Yeoprl96_2006,Liuzzoprl119_2017} and represents a fundamental ingredient to the progress of many quantum technologies such as quantum gate teleportation \cite{Gottesmannature402_1999}, quantum repeaters \cite{Briegelprl81_1998,Sangouardrmp83_2011}, measurement-based quantum computing \cite{Raussendorfprl86_2001}, port-based teleportation \cite{Ishizakaprl101_2008} and quantum network teleportation (QTN) \cite{Landryjosb24_2007,Valivarthinatphot10_2016,Sunnatphot10_2016}. Teleportation has also been used as a quantum simulator for 'extreme' phenomena, such as closed timelike curves and the grandfather paradox \cite{Lloydprl106_2011}.

Quantum teleportation, first proposed by Bennett \textit{et al.} \cite{Bennettprl70_1993}, is a scheme of quantum information processing which allows the transfer of a quantum state between remote physical systems without physical transfer of the information carrier. Specifically, an unknown quantum state of a physical system is measured and subsequently reconstructed at a remote location through the use of classical communication and quantum entanglement \cite{Massarprl74_1993,Popescuprl72_1994}. Without entanglement, such quantum state transfer would not be possible within the laws of quantum mechanics. For that reason, quantum teleportation is thought of as the quantum information protocol which clearly demonstrates the character of quantum entanglement as a resource.

To date, quantum teleportation has been achieved and studied in many different systems, including photonic systems, nuclear mag­netic resonance, optical modes, trapped atoms and solid-state systems (see \cite{Pirandolanatphot9_2015} and references therein). Naturally, most attention has been focused on teleporting the state on long-distance \cite{Yinnature488_2012,Manature489_2012} with the recent satellite-based implementations \cite{Rennature549_2017}. However, even though quantum teleportation is a typically bipartite process, it can be extended to multipartite quantum protocols which have not been thoroughly studied yet. Such multipartite protocols are expected to form fundamental components for larger-scale quantum communication and computation \cite{Nielsencambridge_2000}. 

An important extension of quantum teleportation to a multipartite case is known as the controlled quantum teleportation (CQT) \cite{Karlssonpra58_1998} which allows for remote quantum nondemolition (QND) measurements and forms a backbone of QTN 
\cite{Loockprl84_2000,Yonezawanature431_2004, Pirandolanatphot9_2015}. In the simplest case of tripartite systems, the essential concept of the CQT scheme is that the transfer of the quantum state from sender (Alice) to receiver (Bob) needs controller's (Charlie's) classical information and thus, Charlie can determine success or failure of teleportation by restricting the access to his information, what is commonly thought of as a clear manifestation of tripartite entanglement \cite{Yonezawanature431_2004}. When Alice, Bob and Charlie can choose any one of them to be the sender, receiver and controller, then the CQT protocol is equivalent to QTN, a prelude for a genuine quantum internet \cite{Castelvecchinature554_2018}. Here, it is also believed that parties must share a multipartite entangled state to allow teleportation between any two parties \cite{Yonezawanature431_2004, Pirandolanatphot9_2015}. Furthermore, the CQT protocol as discussed in this paper, may be applied in the processing of quantum secret sharing, a prominent quantum-information protocol \cite{Hillerypra59_1999}. 

Although, several implementation schemes of CQT have been proposed over time using, for instance, a Brown state via cavity QED \cite{Wangoptcomm282_2009}, quantum dots \cite{Wangpra87_2013,Heoscirep7_2017}, GHZ-like states \cite{Naseriqip14_2015}, a GHZ state in an ion-trapped system \cite{Xuijtp55_2016}, so far to the best of our knowledge, the successful experimental realization of the CQT protocol has been reported only for GHZ state of continuous variables \cite{Yonezawanature431_2004}. For such system the GHZ teleportation channel can be contracted, for instance, using three vacuum states in the limit of infinite squeezing \cite{Adessopra73_2006}. Naturally, in a real experiment, a maximally entangled GHZ state of continuous variable is not available because of finite squeezing and inherent losses. Therefore, the realistic state generated by three highly squeezed vacuum states is the non-maximally entangled GHZ-like state. Consequently, CQT of a coherent state was perform with fidelity up to $F_{CQT}=64 \%\pm 2\%$ \cite{Yonezawanature431_2004}. 

To overcome the limitation caused by finite squeezing, in this paper we present the first experimental verification of CQT on GHZ states of discrete variables. Using the four-photon source based on the process of spontaneous parametric down-conversion (SPDC), we generate GHZ state and perform the CQT with fidelity of $F=83.0\%\pm 7.3\%$. Our experiment is also successfully repeated for other teleportation channels based on the GHZ states, in particular, a statistical mixture of such states, demonstrating the controller's capability of steering the teleportation process based on the classical correlations without presence of multipartite entanglement. Such result represents a universal feature of CQT and QTN which is deeply rooted in the operational definition of bipartite entanglement \cite{Barasinskisr8_15209_2018}.

%%%%%%%%%%%%%%%%%%%%%%%%%%%%%%%%%%%%%%%%%%%%%%%%%%%%%%%%%%%%%%%%%%%
%\section{The concept of quantum controlled teleportation}
%%%%%%%%%%%%%%%%%%%%%%%%%%%%%%%%%%%%%%%%%%%%%%%%%%%%%%%%%%%%%%%%%%%

{\em The concept of quantum controlled teleportation.---}We start by reviewing the basic tripartite CQT protocol in finite-dimensional settings \cite{Karlssonpra58_1998}.

The protocol considers three remote parties \textendash~Alice, Bob and Charlie \textendash~who share pure three-qubit entangled state in advance. In the perfect scheme, the shared entangled state is taken to be a maximally entangled GHZ state, $|\mathcal{G}^{(1)}\rangle = \frac{1}{\sqrt{2}} \{|H_1H_2H_3\rangle+|V_1V_2V_3\rangle\}$, where we use the polarization degree of freedom of the photons generated in optical setup with $|H\rangle$ and $|V\rangle$ denoting the horizontal and vertical polarization states, respectively. Initially, Alice is in possession of a qubit in mode no. $1$ of the GHZ state and a single qubit in mode $4$ in the input quantum state $|\psi_4\rangle$ which she wants to teleport. In our experiment, the input state is the polarization of an arbitrary single photon: $|\psi_4\rangle=\alpha |H_4\rangle+\beta |V_4\rangle$, where $|\alpha|^2+|\beta|^2=1$. Suppose now that Alice applies a specific joint quantum measure­ment which projects photons in modes $1$ and $4$ into the maximally entangled Bell state $|\psi^{-}_{14}\rangle = \frac{1}{\sqrt{2}} \{|H_{1}\rangle|V_{4}\rangle-|V_{1}\rangle|H_{4}\rangle\}$. As a result, the state of remaining two qubits is simultaneously projected into $|\psi_{23}\rangle =  \alpha |H_{2}\rangle|V_{3}\rangle-\beta |V_{2}\rangle|H_{3}\rangle$ which can be further decomposed in the new basis $\mathcal{B}_{\pm}=\{|+\rangle,|-\rangle\}$ as $|\psi_{23}\rangle = \frac{1}{\sqrt{2}} \{\alpha |H_{2}\rangle-\beta |V_{2}\rangle\}|+_{3}\rangle - \frac{1}{\sqrt{2}} \{\alpha |H_{2}\rangle+\beta |V_{2}\rangle\}|-_{3}\rangle$, where $|\pm_{3}\rangle=\frac{1}{\sqrt{2}} \{|H_{3}\rangle \pm |V_{3}\rangle\}$. In the next step, Charlie (the controller) applies a von Neumann measurements on qubit in mode $3$ in the basis $\mathcal{B}_{\pm}$. Consequently, the final state of qubit in mode $2$, kept by Bob, is equal to $|\psi_2\rangle=\alpha |H_2\rangle+\beta |V_2\rangle$ up to a unitary operation that depends on the outcomes of Charlie's measurements. In contrary, if Charlie decides to apply a von Neumann measurements on mode $3$ in the basis $\mathcal{B}_{HV}=\{|H\rangle,|V\rangle\}$ then the resulting quantum state of Bob's qubit is either $|\psi_2\rangle= |H_2\rangle$ or $|\psi_2\rangle=|V_2\rangle$. These two scenarios clearly show Charlie's power to determine success and failure of CQT.

Now, it is important to note that the above-mentioned $|\psi^{-}_{14}\rangle$ state is only one of four possible Bell states which can be obtained by Alice. In general, the composed state of qubits in modes $1$ and $4$ can be projected into four different states $(\mathcal{P}_k \otimes I)|\psi^{-}_{14}\rangle$, where $\mathcal{P}_k$ is an appropriate Pauli operator \cite{Nielsencambridge_2000} and $k=0,1,2,3$. When this happens, the state of particles in modes $2$ and $3$ becomes $\rho_{23}=\mathcal{P}^{\dagger}_k |\psi_{23}\rangle \langle \psi_{23}| \mathcal{P}_k$. Bob can then recover the input state by applying an accordingly chosen transformation that requires a classical communication with both Alice and Charlie. Although in the above-mentioned scheme only one of four Bell states is distinguished, teleportation is still successfully achieved, albeit only in a quarter of the cases. Moreover, it should be noted that the complete Bell state measurement which is based on nonlinear processes, requires hyperentanglement or feed-forward techniques \cite{Knillnature409_2001} and hence, it remains an experimentally challenging problem which usually causes the reduction of the signal intensity \cite{Kimprl86_2001,Gricepra84_2011}. Therefore, the antisymmetric structure of the state $|\psi^{-}_{14}\rangle$ makes this state the most useful in the experimental implementation of teleportation protocols as discussed in \cite{Bouwmeesternature390_1997,Kimprl86_2001}. In this paper, we also take the advantage of this property and limit our Bell state measurement only to $|\psi^{-}_{14}\rangle$.

Finally, we note that the faithfulness of the CQT protocol shall not change if one applies a local bit flip operation on the GHZ state shared by Alice, Bob and Charlie, say  $|\mathcal{G}^{(2)}\rangle = \frac{1}{\sqrt{2}} \{|H_1H_2V_3\rangle+|V_1V_2H_3\rangle\}$. A particularly interesting scenario, however, occurs if one takes a statistical mixture of such two GHZ states, 
\begin{equation}
\rho(p)= (1-p) |\mathcal{G}^{(1)}\rangle\langle \mathcal{G}^{(1)}|+ p |\mathcal{G}^{(2)}\rangle\langle \mathcal{G}^{(2)}|,
\label{eq:GHZ_mixture}
\end{equation}
where $0\leq p \leq 1$. Then, for the equivalently balanced probabilities the state $\rho(p=1/2)$ belongs to the biseparable class and can be decomposed as $\rho(p=1/2)=\frac{1}{2} \{|\chi^{+}\rangle\langle\chi^{+}|+|\chi^{-}\rangle\langle\chi^{-}|\}$, where $|\chi^{\pm}\rangle = \frac{1}{2} \{|H_1H_2\rangle \pm |V_1 V_2\rangle\} \otimes \{\pm|H_3\rangle + |V_3\rangle\}$. This means that there are no other correlations between Charlie and the rest of the system besides the classical ones. Despite that Charlie's capability of controlling the teleportation protocol remains unchanged 
\cite{Barasinskisr8_15209_2018}. 

To emphasize the significant role of tripartite (fully-entangled and biseparable) states in the CQT protocol, let us discuss the difference between CQT and the classical control of ordinary teleportation. Suppose that Alice and Bob share either of two Bell states $|\phi^{+}_{12}\rangle = \frac{1}{\sqrt{2}} \{|H_1H_2\rangle + |V_1V_2\rangle\}$ or $|\phi^{-}_{12}\rangle = \frac{1}{\sqrt{2}} \{|H_1H_2\rangle - |V_1V_2\rangle\}$ with equal probability and the information which Bell state is truly shared belongs only to Charlie. Then, the teleportation between Alice and Bob is successfully performed when Charlie broadcasts the information he has and is forbidden otherwise. 
Analogously, for the quantum protocol the GHZ state can be written in the form $|\mathcal{G}^{(1)}\rangle = \frac{1}{\sqrt{2}} (|\phi^{+}_{12}\rangle |+_3\rangle + |\phi^{-}_{12}\rangle |-_3\rangle )$, where the information “which Bell state” is encoded in the Charlie’s qubit via an QND-type interaction, i.e. in the basis of $|+_3\rangle$ and $|-_3\rangle$. 
However, in this case the information is quantum-mechanically possessed by Charlie and hence, any measurement in the logical basis implies that the teleportation is
forbidden principially and not just by Bob's ignorance of Charlie's
outcome (e.g. it can not be restored by any eyevesdropping).
This is the main difference with the classical counterpart, valid even for the biseparable mixture $\rho(p=1/2)$. 
Note that measurement in any other basis than $|0/1\rangle$ allows to restore the teleportation at least probabilistically by implementing an appropriate filtering. 
%%%%%%%%%%%%%%%%%%%%%%%%%%%%%%%%%%%%%%%%%%%%%%%%%%%%%%%%%%%%%%%%%%%
%\section{Experimental implementation}
%%%%%%%%%%%%%%%%%%%%%%%%%%%%%%%%%%%%%%%%%%%%%%%%%%%%%%%%%%%%%%%%%%%

\begin{figure}
\centering
\includegraphics[width=\linewidth]{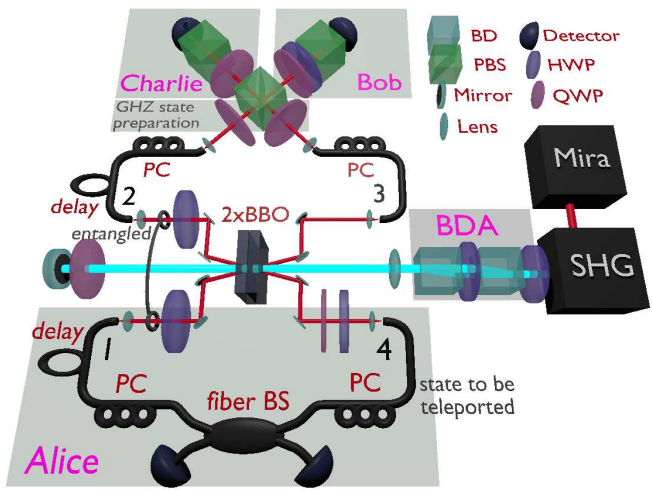}
\caption{Scheme of the experimental setup for the controlled quantum teleportation as described in the
text. The components are labeled as follows: BS \textendash~ beamsplitter, PBS \textendash~ polarizing beamsplitter, PC \textendash~ polarization controller, HWP \textendash~ half-waveplate, QWP \textendash~  quarter-waveplate, BDA \textendash~ beam displacer assembly, BD \textendash~ beam displacer. The abbreviation SHG stands for the second-harmonic generation and Mira is the femtosecond laser system manufactured by Coherent.
}
\label{fig:fig2}
\end{figure}

{\em Experimental implementation.---}The experimental setup consists of a four-photon source, GHZ preparation stage and three stations operated by Alice, Bob and Charlie (see Fig. \ref{fig:fig2}). Photons are generated in a BBO crystal cascade \cite{Kwiat} by means of spontaneous parametric down-conversion pumped by femto-second pulses at $413$\,nm. First pair of photons (modes $1$ and $2$) is generated while the pumping pulse propagates through the crystals in the forward direction. Subsequently it gets reflected on a mirror and generates a second pair of photons (modes $3$ and $4$) on its way back. Pump beam polarization is controlled by a half-wave plate (HWP) and a polarization dispersion line (BDA) to correct for polarization group velocity dispersion \cite{NambuPRA66}. Proper polarization of the pumping beam allows to generate photons in modes $1$ and $2$ in an entangled state $|\Phi^+_{12}\rangle = \frac{1}{\sqrt{2}} \{|H_1 H_2\rangle - i|V_1 V_2\rangle\}$. Photons $3$ and $4$ are collected only from one of the crystals obtaining thus a separable state $|H_3 H_4\rangle$. After being subjected to procedures described below, the photons are collected to single-mode optical fibers and detected by a set of four avalanche photodetectors. Simultaneous four-fold coincidence detections are recorded.

In the next step, we generate the GHZ state $|\mathcal{G}^{(1)}\rangle$ in Eq. \eqref{eq:GHZ_mixture}.
To achieve that, polarization of the photon in mode $3$ is changed to circular $|R_3\rangle=\frac{1}{\sqrt{2}}\{|H_3\rangle + i|V_3\rangle\}$ and then it overlaps with photon in mode $2$ on the polarizing beamsplitter $PBS$. With success probability of $\frac{1}{2}$, these two photons leave $PBS$ by different output ports and together with the photon in mode $1$ form the GHZ state $|\mathcal{G}^{(1)}\rangle$ \cite{Lupra78_2008}. Although we have not repeated the GHZ state preparation testing in the same way as the authors of Ref. \cite{Bussieresnatutephot8_2014}, we have performed testing of individual component blocks of our setup, namely we have observed purities of about $90\%$ of the generated Bell state (modes $1$ and $2$) together with similarly pure Bell state preparation with photons in mode $2$ and $3$ on the $PBS$ (note: this was tested when projecting mode $1$ onto the $|H\rangle$ state). We label the photon modes leading to Bob's and Charlie's apparatus by nos. $2$ and $3$ respectively.

Alice subsequently encodes the to-be-teleported qubit into the polarization state of the photon in mode $4$ using  a HWP and QWP. Then she projects the state of photons in modes $1$ and $4$ onto a singlet Bell state $|\psi^{-}_{14}\rangle$ by post-selecting on photon anti-bunching behind a balanced fiber beamsplitter.

At this point, Charlie decides whether to allow or deny the teleportation. In order to allow it, Charlie projects the state of the photon in mode $3$ to circular polarization. Similarly, to deny the teleportation, Charlie projects his photon onto horizontal polarization. Due to the nature of coincidence-based measurement, Alice's and Charlie's actions happen simultaneously.

Bob receives the teleported qubit encoded in the state of the photon in mode $2$. He then subjects this photon to a polarization projection measurement using a sequence of HWP, QWP and a polarizer. 
To evaluate the performance of teleportation, we measure the fidelity of the tele­ported state $F =  \langle \psi_4 | \rho_2 |\psi_4\rangle$, where $\rho_2$ is the resulting state of photon in mode $2$ (mixed in general).  
Based on the coincidence counts observed for different combinations of input states encoded by Alice and Bob's projection measurement, $F$ is calculated as \cite{boschiprl1998}
\begin{equation}
F =  \frac{f_{\|}}{f_{\|}+f_\perp},
\end{equation}
where $f_{\|}$ stands for coincidence rate observed when Bob projects on the state identical to Alice's encoding choice. Likewise, $f_\perp$ stands for coincidence rate observed when Bob projects on an orthogonal state.

%%%%%%%%%%%%%%%%%%%%%%%%%%%%%%%%%%%%%%%%%%%%%%%%%%%%%%%%%%%%%%%%%%%
%\section{Experimental results}
%%%%%%%%%%%%%%%%%%%%%%%%%%%%%%%%%%%%%%%%%%%%%%%%%%%%%%%%%%%%%%%%%%%
\begin{table}%[htbp]
\centering
\begin{tabular}{ccc}
\hline
channel$ \quad$ & $F_{\textrm{allowed}} (\%)\quad$ & $F_{\textrm{denied}} (\%)\quad$ \\
\hline
$\rho_{\textrm{ref}} \quad$	& $83.1 \pm 4.9 \quad$ & -- \\
$\rho(p=0) \quad$ & $77.9 \pm 8.1 \quad$ & $57.2 \pm 5.0 \quad$ \\
$\rho(p=1) \quad$ & $83.0 \pm 7.5 \quad$ & $51.8 \pm 6.7 \quad$ \\
$\rho(p=1/2) \quad$ & $80.2 \pm 5.7 \quad$ & $55.1 \pm 5.0 \quad$ \\
\hline
\end{tabular}
\caption{Measured fidelities for the linearly polarized input state $|\psi_4\rangle=\frac{1}{\sqrt{2}} \{|H_4\rangle+|V_4\rangle\}$ and several teleportation channels: $\rho_{\textrm{ref}}$ denotes two-photon teleportation channel $|\Phi^+_{12}\rangle\langle \Phi^+_{12}|$ with photon in mode  $3$ serving only as trigger (see text) while $\rho(p)$ is given in Eq. \eqref{eq:GHZ_mixture}. The last two columns correspond to the process when controller allows and denies the teleportation, respectively. All uncertainties are obtain by numerical calculations assuming a Poisson distribution of the four-fold coincidences.
}
\label{tab:shape_functions}
\end{table}

{\em Experimental results.---}We test the CQT protocol on a linearly polarized balanced ($\alpha=\beta$) input state i.e. $|\psi_4\rangle=\frac{1}{\sqrt{2}} \{|H_4\rangle+|V_4\rangle\}$. This choice of $|\psi_4\rangle$ is quite natural i.e. by the very description of the GHZ state preparation as both $|H\rangle$ and $|V\rangle$ polarization can be considered as preferred directions in the experiment. Therefore, the input state polarized at $45^{\circ}$ represents one of the most challenging tasks and other commonly analyzed states yield the teleportation fidelity approximately equal or greater \cite{Bouwmeesternature390_1997, Bussieresnatutephot8_2014, Pfaffscience345_2014, Rennature549_2017}.

For the purposes of preliminary testing, in the first experiment we have operated our setup in the regime of ordinary uncontrolled quantum teleportation \cite{Bennettprl70_1993}. To achieve that, polarization of the third photon is kept horizontal to be directly transmitted on $PBS$ on to Charlie's detector. As for Bob's projection measurement, it is performed using the combination of $HWP$ and $PBS$. Again, four-fold coincidences are registered, this time with photon in mode 3 serving only as a trigger. Based on Eq. \eqref{eq:GHZ_mixture} the faithfulness of uncontrolled quantum teleportation has been found to be $F=83.1\% \pm 4.9 \%$ which is in line with recent experiments on photonic qubits (e.g. \cite{Bussieresnatutephot8_2014,Rennature549_2017}).

Probabilistic nature of our four-photon source causes undesired higher-order SPDC terms to contribute to the detected signal. Presented fidelities therefore need to be corrected for these imperfections of the source to be faithful characteristics of the protocol implementation itself. A detailed analysis of these corrections is presented in the Supplementary material \cite{supplement}.

\begin{figure}%[htbp]
\centering
\includegraphics[width=\linewidth]{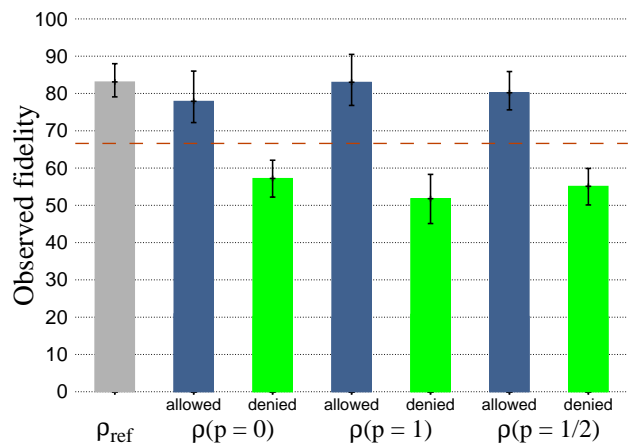}
\caption{Teleportation fidelities measured for several quantum channels (see Table \ref{tab:shape_functions}). Gray bar refers to a standard uncontrolled quantum teleportation performed as a preliminary test. Blue bars correspond to teleportation faithfulness $F_{\textrm{allowed}}$ achieved with controller's permission while $F_{\textrm{denied}}$ are shown in green. The horizontal dashed line marks the classical limit of $66.7\%$. Black line segments represent the confidence intervals.}
\label{fig:fig3}
\end{figure}

In the second experiment, we have performed the teleportation on the GHZ state. In order to do this, we set back the polarization of the photon in mode $3$ to circular, $|R_3\rangle$, thus generating the $|\mathcal{G}^{(1)}\rangle$ channel. By proper operating of the QWP and the polarizer we analyze two scenarios of teleportation. In the first one, when Charlie allows for the teleportation the fidelity calculated from the four-fold coincidences $F_{\textrm{allowed}}=77.9\% \pm 8.1 \%$. This result exceeds the classical limit of $66.7\%$ and thus certifies  the quantum nature of our teleportation experiment. In the second scenario, i.e. without Charlie's permission, the fidelity of $F_{\textrm{denied}}=57.2 \pm 5.0 \%$ meets the second condition of CQT. This results clearly shows the success of CQT through the GHZ state of discrete variables. Similar measurements  have been performed for $|\mathcal{G}^{(2)}\rangle$ channel. This kind of GHZ state can be prepared by slight modification of the experimental setup. Specifically, both HWPs in the mode 2 and the Bob's analyzer part, are rotated by $\pi/4$. The corresponding fidelities are presented in Table \ref{tab:shape_functions} and visualized in Fig. \ref{fig:fig3}. What is important, in this configuration the CQT is realized with even greater fidelity of around $83\%$ with simultaneous decrease of $F_{\textrm{denied}}$.

Now we perform the CQT through the statistical mixture $\rho(p)$ given in Eq. \eqref{eq:GHZ_mixture} when $p=1/2$.
To emulate this, we have simply summed up the respective coincidence counts obtained for $|\mathcal{G}^{(1)}\rangle$ and $|\mathcal{G}^{(2)}\rangle$.
We found the resulting fidelity to be $F_{\textrm{allowed}}=80.2\% \pm 5.7 \%$ when Charlie permits the teleportation and $F_{\textrm{denied}}= 55.1\% \pm 5.0 \%$ otherwise. This means that both conditions of CQT are satisfied also for $\rho(p=1/2)$ channel, despite it belongs to the biseparable class and there is no entanglement between Charlie's photon and the remaining two photons. 
In order to verify this fact experimentally one can use the well know methods such as $\theta-$protocol \cite{McCutcheonnaturecommun7_2016} or $XY-$protocol \cite{Pappaprl108_2012}. In fact, both protocols have been successfully applied recently for experimental detection of tripartite entanglement in the GHZ states \cite{McCutcheonnaturecommun7_2016}. Since we use the entangled-photon source with the same efficiency as in Ref. \cite{McCutcheonnaturecommun7_2016} the outcome of multipartite-entanglement-detection protocol is similar and it is out of the scope of this paper. To the best of our knowledge, this is the first demonstration of CQT based on the biseparable states.

%%%%%%%%%%%%%%%%%%%%%%%%%%%%%%%%%%%%%%%%%%%%%%%%%%%%%%%%%%%%%%%%%%%
%\section{Conclusions}
%%%%%%%%%%%%%%%%%%%%%%%%%%%%%%%%%%%%%%%%%%%%%%%%%%%%%%%%%%%%%%%%%%%

{\em Conclusions.---}In summary, we have presented a proof-of-principle experimental demonstration of CQT through various kinds of GHZ states of discrete variables with the fidelities well above the classical limit. 
Our experiment shows that tripartite entanglement is not a necessary recourse for CQT. In fact, the classical correlation between controller and joined "sender-receiver" subsystem is sufficient in order to allow or forbid the teleportation. 
In a broader context, our results open new possible ways of implementation of CQT lowering requirements for a state preparation and preservation what is of practical importance in realizing more complicated quantum computation and quantum communications among many parties. 
In particular, one can consider the three-qubit Werner state ($\rho_W = q |\mathcal{G}^{(1)}\rangle \langle \mathcal{G}^{(1)}| + (1-q) I/8$) which can be thought of as an imperfect preparation of the GHZ quantum channel. 
The ability to perform CQT through biseparable Werner states (i.e. for $1/3 < q \leq 3/7$) implies that the CQT is less fragile against noise than the tripartite entanglement, as described in \cite{Barasinskisr8_15209_2018}. 
Furthermore, as the three-qubit Werner states are invariant under qubits permutation, the CQT can be successfully performed no matter how we split the qubits between Alice, Bob and Charlie. In our experiment, despite the probabilistic nature of the GHZ state preparation and the teleportation itself, the roles of Alice, Bob and Charlie can also be swapped (see Supplementary material for detailed analysis \cite{supplement}). 
In other words, fundamentals of our experiment can be easily used in demonstration of a QTN for biseparable states. This conclusion is in contrast with common opinion "Only if we use a fully inseparable tripartite entangled state can we succeed in teleportation between an arbitrary pair in the network" \cite{Yonezawanature431_2004}. 
Explanation of this phenomenon is based on the concept of localizable entanglement \cite{Verstraeteprl92_2004} which plays a central role in CQT and QTN \cite{Barasinskisr8_15209_2018}. Our experiment shows non-trivial application of localizable entanglement leading to results which cannot be predicted by standard quantifiers of multipartite entanglement.

%%%%%%%%%%%%%%%%%%%%%%%%%%%%%%%%%%%%%%%%%%%%%%%%%%%%%%%%%%%%%%%%%%%
%\section{Acknowledgments}
%%%%%%%%%%%%%%%%%%%%%%%%%%%%%%%%%%%%%%%%%%%%%%%%%%%%%%%%%%%%%%%%%%%

\begin{acknowledgments}
A.B. was supported by GA \v{C}R Project No. 17-23005Y, A.Č. and K.L. by Project No. 17-10003S. Authors also thank M\v{S}MT \v{C}R for support by the project CZ.$02.1.01/0.0/0.0/16\_019/0000754$ and Cesnet for providing data management services.
\end{acknowledgments}

\end{document}

\title{Demonstration of quantum controlled teleportation for discrete variables on linear optical devices -- Supplementary material} 
\author{Artur Barasi\'nski}
\email{artur.barasinski@upol.cz}
\affiliation{RCPTM, Joint Laboratory of Optics of Palack\'{y} University and Institute of Physics of CAS, Faculty of Science, Palack\'{y} University, 17. Listopadu 12, 771 46 Olomouc, Czech Republic}
\affiliation{Institute of Physics, University of Zielona G\'ora, Z. Szafrana 4a, 65-516 Zielona G\' ora, Poland}

\author{Anton\'{i}n \v{C}ernoch}
\email{antonin.cernoch@upol.cz}
\affiliation{RCPTM, Joint Laboratory of Optics of Palack\'{y} University and Institute of Physics of CAS, Faculty of Science, Palack\'{y} University, 17. Listopadu 12, 771 46 Olomouc, Czech Republic}

\author{Karel Lemr}
\email{k.lemr@upol.cz}
\affiliation{RCPTM, Joint Laboratory of Optics of Palack\'{y} University and Institute of Physics of CAS, Faculty of Science, Palack\'{y} University, 17. Listopadu 12, 771 46 Olomouc, Czech Republic}

\maketitle

\section{Effect of source imperfection on protocol fidelity}
In this experiment, photons were obtained using the process of spontaneous parametric down-conversion (SPDC). This non-linear optical process is described by the effective interaction Hamiltonian
\begin{equation}
\hat{H}_\mathrm{eff} = \kappa \hat{a}^\dagger_s \hat{a}^\dagger_i,
\end{equation}
where $\hat{a}^\dagger_{s,i}$ denote creation operators in the signal and idler modes respectively. $\kappa$ is an overall interaction strength incorporating all material parameters as well as the power of pumping laser beam in the undepleatable approximation \cite{boyd,rubin}. In our configuration, it holds that $|\kappa| \ll 1$. Assuming both signal and idler modes to be in a vacuum state prior to the interaction, their output state can be expressed in the Fock basis as
\begin{equation}
\label{eq:SPDCout}
|\psi\rangle_\mathrm{SPDCout} \propto |00\rangle + \kappa |11\rangle + \kappa^2 |22\rangle + \kappa |33\rangle+...
\end{equation}
with numbers denoting the number of photons in signal and idler modes. To create four photons, each in its respective mode, we implement the SPDC process twice, pumping the crystal in the forward and backward direction as explained in the main text. The resulting four-mode Hamiltonian can be described as sum of two single-SPDC Hamiltonians $\hat{H}_\mathrm{eff}$. The overall four-mode output state is, thus, in the form of
\begin{eqnarray}
\label{eq:totalout}
|\psi\rangle_\mathrm{tot} & \propto & |0000\rangle + \kappa |0011\rangle + \kappa |1100\rangle + \kappa^2 |1111\rangle + \nonumber\\
& + & \kappa^2 |2200\rangle + \kappa^2 |0022\rangle + ...,
\end{eqnarray}
where terms containing $\kappa$ in more then second power were omitted as they are negligible in the $|\kappa| \ll 1$ approximation. Similarly, we can immediately disregard the first three terms as they can not produce a four-photon coincident detection used to post-select on succesful events. Note that detector dark-counts also represent a negligible effect. The relevant three terms that contribute to the detected signal consist of the desired term $|1111\rangle$ (one photon in each mode) and the two undesired terms $|0022\rangle$ and $|2200\rangle$ (two modes populated by two-photon state each, vacuum elsewhere). Due to the probabilistic nature of the SPDC process one can not distinguish detection events caused by either of these three terms and one, thus, relies on subtraction of the undesired coincidences (caused by $|0022\rangle$ and $|2200\rangle$) during signal post-processing. Also note that in Eq. (\ref{eq:totalout}), we assume identical interaction strength for both SPDC processes. This is not the case in reality as the pumping in backward direction becomes weaker becuse of longer propagation path and additional reflections on mirrors (or at least one mirror). The presented analysis also does not discuss the entanglement generation in a BBO crystal cascade to keep it simple within confines its purpose. The reader is ecouraged to consult Ref. \cite{bovino} for a more detailed account.

In order to implement correction for the undesired SPDC terms, we have measured the two-photon coincident detections in the forward and backward SPDC modes to asses the relative interaction strengths of the two SPDC processes. Subsequently, we have calculated the theoretical success probability (probability of detecting a four-fold coincidence) considering each of the three relevant input states (i.e. $|1111\rangle$, $|0022\rangle$, and $|2200\rangle$) propagating through the setup. Note that we have not corrected for any other setup imperfections except the imperfect polarizing beam splitter used in GHZ state preparation (reflecting about 5 \% of horizontally polarized light). Based on this analysis and the relative strength of forward and backward SPDC processes, we have estimated the percentage of undesired four-fold coincidences for the cases of uncontrolled, controlled allowed, and controlled denied teleportation. The respective values read: 13.0 \%, 55.4 \%, and 30.1 \%. We have also identified that in case of undesired terms, Bob receives an output qubits completely uncorrelated with Alice's to-be-teleported input state. Bob simply observes a maximally mixed state $\hat{I}/2$. Using this fact, we have performed the correction on undesired terms in the following way: raw coincidence data were used to estimate Bob's output qubit density matrix using a maximum likelihood method \cite{jezek}. The obtained density matrix was then corrected by subtracting a mixed state with a  weight corresponding to the percentage of undesired coincidences. The resulting density matrix was renormalized and used for fidelity estimation.

Without any correction for undesired SPDC terms, the observed fidelities would be significantly lower. We present their values in Table \ref{tab:fidel}. The reader shall however be aware that the goal of this experiment was to demonstrate a controlled-teleportation protocol with discrete variables. The imperfections of the source are a separate issue and shall not be directly interpreted as imperfections of the protocol implementation itself.
\begin{table}
\caption{\label{tab:fidel} Measured raw fidelities without corrections for undesired photon-number terms. For the meaning of respective channels, see the main text.}
\begin{ruledtabular}
\begin{tabular}{lll}
Channel & $F_\mathrm{allowed}$ (\%) & $F_\mathrm{denied} (\%)$\\\hline
$\hat{\rho}_\mathrm{ref}$ & $78.8 \pm 4.7$& N/A\\
$\hat{\rho}(p=0)$ & $62.4 \pm 1.6$& $55.0 \pm 1.4$\\
$\hat{\rho}(p=1)$ & $64.7 \pm 1.9$& $51.2 \pm 0.9$\\
$\hat{\rho}(p=1/2)$ & $63.5 \pm 1.2$& $53.5 \pm 0.9$
\end{tabular}
\end{ruledtabular}
\end{table}
\section{Switching the roles of Alice, Bob, and Charlie}
In the presented scheme, Alice teleports a qubit state to Bob while Charlie decides on either allowing or denying the process to happen. The general quantum networks require that the roles of the respective parties can be swapped (e.g. Bob teleports a qubit to Charlie under Alice's control). In order to do so, the probabilistic GHZ state preparation would need to be followed by a singlet state projection, for instance, at Bob's end of the channel. On probabilistic platforms, such as the platform of linear optics, it is often impossible to simply chain two probabilistic gates as the required post-selection on both of them are in conflict.

Quite remarkably, in this case, the post-selection on correct GHZ state preparation and post-selection on singlet state are not in mutual conflict assuming a perfect four-photon source (each photon in its proper mode). In our setup, the GHZ state is prepared by post-selecting on coincident detection by Bob and Charlie. Let us now assume that it is Bob who wishes to teleport a qubit to Charlie (Alice is the controller). Bob needs to implement a singlet projection on his output from the GHZ state preparation stage and on the teleported photon. That is, he needs to add a beam splitter to overlap the teleported photon with his output of the GHZ state preparation. Two photons enter the GHZ preparation implemented by the polarizing beam splitter, each by one of its input ports. Bob's output might therefore contain zero, one or two photons. Only if one photon leaves by Bob's output and the other by Charlie's, the GHZ state is successfully prepared. Now assume that there are no photons at Bob's GHZ preparation output. Such vacuum state together with the teleported photon can not produce a two-fold coincident detection heralding a singlet state projection and therefore would not contribute to the observed results. Similarly, if there are two photons at Bob's GHZ state output, Charlie can not detect any photon and the protocol can therefore not succeed. The fact that post-selections on GHZ state preparation and singlet projections are not in conflict allows to chain these two operations even though they are probabilistic. As a result, one can swap the roles of Alice, Bob and Charlie and implement a fully-featured elementary quantum network.

One will be able to swap the roles of the three parties even when an imperfect four-photon source is used (as described in the previous section). It would, however, be required to perform similar corrections as described in the previous section, but with different weights, to compensate for undesired multi-photon terms.